\documentclass[twocolumn,showpacs,prl]{revtex4}
\usepackage{graphicx}

\def\ip{in-plane}
\def\ea{\emph{et al.}}
\def\ie{\emph{i.e.}}

\def\aif{$\alpha_{i}$}
\def\msgd{$M_s^{Gd}$}
\def\msfe{$M_s^{Fe}$}

\def\Aint{$A_{int}$}
\def\dfe{$\delta_{Fe}$}
\def\dgd{$\delta_{Gd}$}
\def\tfe{$t_{Fe}$}
\def\tgd{$t_{Gd}$}

\def\fcgd{$\phi_c^{Gd}$}

\begin{document}

\title{Ground state and constrained domain walls in Gd/Fe multilayers}

\author{Bas B. \surname{Van Aken}}\email{aken@mbi-berlin.de}
\altaffiliation[Present address: ]{Max-Born-Institut, Berlin,
Germany.}\affiliation{Department of Materials
Science and Metallurgy, University of Cambridge, Cambridge, CB2
3QZ, UK}
\author{Jos\'{e} L. Prieto}\altaffiliation[Present address: ]
{Dep. Fisica Aplicada, Universidad Polit\'{e}cnica de Madrid,
Madrid, Spain.}\affiliation{Department of Materials Science and
Metallurgy, University of Cambridge, Cambridge, CB2 3QZ, UK}
\author{Neil D. Mathur}\affiliation{Department of Materials
Science and Metallurgy, University of Cambridge, Cambridge, CB2
3QZ, UK}

\begin{abstract}
The magnetic ground state of antiferromagnetically coupled Gd/Fe
multilayers and the evolution of \ip\ domain walls is modelled
with micromagnetics. The twisted state is characterised by a rapid
decrease of the interface angle with increasing magnetic field. We
found that for certain ratios $M^{Fe} : M^{Gd}$, the twisted state
is already present at low fields. However, the magnetic ground
state is not only determined by the ratio $M^{Fe} : M^{Gd}$ but
also by the thicknesses of the layers, that is the total moments
of the layer. The dependence of the magnetic ground state is
explained by the amount of overlap of the domain walls at the
interface. Thicker layers suppress the Fe aligned and the Gd
aligned state in favour of the twisted state. Whereas ultrathin
layers exclude the twisted state, since wider domain walls can not
form in these ultrathin layers.
\end{abstract}

\maketitle

The study of spin-polarised transport is key for the understanding
of processes involved in spin-electronics. A particular
interesting point is the effect of domain walls (DWs) on the
transport properties of ferromagnetic materials, as has been shown
recently.\cite{Pri03b,Pri04a} This interest is based on the
possibility of creating sharp boundaries (high angular rotation of
the magnetisation in only few atomic layers) without any
non-magnetic spacer.

DWs have been studied in several systems. Domain walls can be
measured in single crystal highly anisotropic thin films of
SrRuO$_3$.\cite{Kle00} Or between features in a patterned
La$_{0.7}$Ca$_{0.3}$MnO$_3$ thin film.\cite{Mat99} Another form of
magnetic boundary has been observed in single crystalline thin
films of Fe$_3$O$_4$, where at stacking faults the coupling is
antiferromagnetic and $180^\circ$ magnetic boundaries are
formed.\cite{Eer02b} It has been shown that domain walls can be
formed in Gd/FM multilayer devices, where FM is a standard
ferromagnetic metal like iron, cobalt or
permalloy.\cite{Cam87a,Che91a} Recently, magnetoresistance
measurements have been reported with current perpendicular to the
in-plane DWs.\cite{Pri03b,Pri04a}

From the point of view of transport properties of FM metals, Gd/FM
multilayers are of special interest for three main reasons:
firstly the Gd/FM interface is an antiferromagnetic (AFM) sharp
boundary, so it is ideal for the study of spin diffusion and
spin-flip. Secondly, there is a large resistance mismatch across
the interface, making the system similar to metal-semiconductor
interfaces. Finally, in certain conditions as explained below,
Gd/FM multilayers can support in-plane domain walls (DWs) close to
the interface, so their transport properties can be studied in
detail.

Current perpendicular-to-plane magnetoresistance experiments in
Gd/Fe multilayers have shown that the resistance decreases with
increased applied field due to a decrease in the interface angle
\aif.\cite{Pri04a} This is in qualitative agreement with
magnetoresistance experiments in Fe$_3$O$_4$, where the resistance
is a function of \aif\ controlled by the applied external
field\cite{Eer02b}. However, one of the samples in Ref.
[\onlinecite{Pri04a}] showed an upturn in the resistance at large
fields and it was speculated that this is related to the formation
of DWs.

The ground state domain walls at the Gd/Fe boundary have been
magnetically characterised in bilayers\cite{Mcg96a} and
multilayers too, but normally in quite thin layers. In these
situations the interface and the AFM coupling play the most
important roles in determining the magnetic structure.
Traditionally, the Fe aligned state is found for $M^{Fe}>M^{Gd}$
and the Gd aligned state when $M^{Fe}<M^{Gd}$. Upon application of
an external field $H$ the twisted state with DWs at the interfaces
is observed. Note the distinction between \msgd\ and $M^{Gd}$.
\msgd\ is the saturation magnetisation per volume and $M^{Gd}$ is
the total moment of all Gd layers together which is thickness
dependent. Most articles focus on the effect of changing ratio
\msgd : \msfe\ with temperature, and therefore changing ratio
$M^{Fe}: M^{Gd}$. Since the Curie temperature of Gd is much
smaller than that of Fe, \msgd\ decreases rapidly with respect to
\msfe\ with increasing temperature.

Micromagnetic simulation has been done to gain insight in the
changes in the magnetic state with applied field. We study the
effect of the thicknesses \tfe\ and \tgd\ on the magnetic
behaviour of Gd/Fe multilayers. We will show that there are
distinct differences between samples with $M^{Fe}\gg M^{Gd}$ and
samples with $M^{Fe}\approx M^{Gd}$. We will also show that at
constant ratio $M^{Fe}: M^{Gd}$, and therefore the magnetic ground
state is significantly influenced by the actual value of \tgd\ and
\tfe. We will present the phase diagram of the magnetic ground
state as a function of \tgd\ and \tfe. In ultra thin films the DWs
on both side of the layer are not independent and interact via
their surface tension. The Fe and Gd aligned state are completely
suppressed for multilayers with \tgd~$>16$ nm and \tfe~$>9$ nm.

In this paper we model Gd/Fe multilayers using the LLG
micromagnetics software.\cite{llg} The modelled multilayers
consist of three Fe layers, thickness \tfe, and two Gd layers with
thickness \tgd. The simulation volume is $10\times 10\times
t_{tot}$ nm$^3$, where $t_{tot}$ is the total thickness of the
multilayer given by $t_{tot}=3$ \tfe\ + 2 \tgd. The cell size in
the simulation is $1\times 1\times 0.25$ nm$^3$, ensuring that the
typical DWs found in this work are at least five to ten cells
thick. In-plane magnetisation is obtained by periodic boundaries
on the $xz$ and $yz$ planes. The materials parameters for Fe and
Gd that have been used are listed in Table~\ref{Tab:param},
effectively the simulations were done at $T=0$ K. The value used
for \msgd\ is 25\% smaller than the bulk value as reported in many
experimental works.\cite{Ish99a,Hah95a,Hos02a} We will discuss the
effect of reduced \msgd\ later on. The AFM Fe-Gd interface
interaction is chosen to be \Aint\ $=-1.0~\mu$erg cm$^{-1}$,
between the exchange coupling strengths of Fe and Gd. Furthermore,
comparing work on the temperature dependence of the Co-Gd coupling
and the Fe-Gd coupling indicates that \Aint\ $=-1.0~\mu$erg
cm$^{-1}$ is a good estimate.\cite{Wuc97a}

\begin{table}[htb]\begin{centering}
\caption{Material parameters for Fe and Gd.} \label{Tab:param}
\begin{tabular}{rll}
 & Fe & Gd \\
 $M_s$ (emu cm$^{-3}$) & 1700 & 1508 \\
 $A$ ($\mu$erg cm$^{-1}$) & 2.05 & 0.75 \\
 $K$ ($\mu$erg cm$^{-1}$) & 1.4$\times 10^5$ & $10^2$ \\
 & (cubic) & (uniaxial) \\
\end{tabular}
\end{centering}
\end{table}

For each simulation, a part of the hysteresis loop has been
modelled from $\mu_0 H=7.5$~T to $\mu_0 H=0$ T. This prevented the
problem of having to choose the correct initial state at $\mu_0
H=0$ T. The initial magnetisation was aligned with the $x$-axis
and the Fe (Gd) magnetisation was parallel (antiparallel) with the
positive direction. The field was applied in the positive $y$
direction. The initial field $\mu_0 H=7.5$ T rotates the bulk of
the layers parallel with the field; the Fe (Gd) boundary
magnetisation however remains a component parallel to the positive
(negative) $x$ axis. All magnetisation vectors within a single
$xy$ simulation layer are identical and the $z$ component is
always negligible ($<10^{-4} M_s$). The angle between the
magnetisation at layer depth $t$ and the applied field direction
is $\phi(t)$. Thus $\phi=0^{\circ}$ corresponds to the
magnetisation aligned with the $y$-axis and $\phi=\pm 90^\circ$ is
aligned with the $x$-axis.

\begin{figure}[htb]
\begin{centering}
\includegraphics[bb=97 405 339 582,width=70mm]{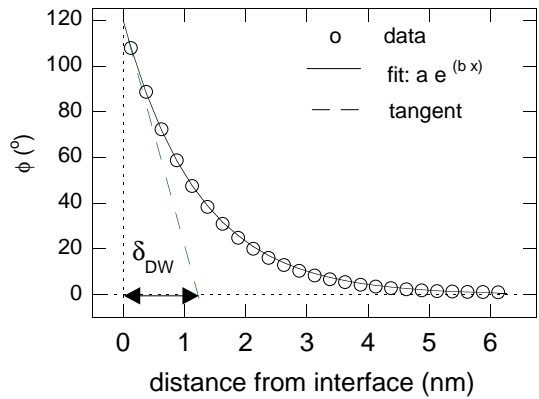}
\caption{{Calculation of the DW width from the calculated
magnetisation angle $\phi$ versus layer depth $t$ data. The drawn
line is a fit to the data ($\circ$). The dashed line is the
tangent at $t'=0$, see text for explanation. The horizontal arrow
indicates the DW width.}} \label{fig:DWs}
\end{centering}
\end{figure}

At low temperatures and zero field, each layer is expected to be
fully FM aligned, and antiparallel with respect to the adjacent
layers, creating an interface angle \aif\ = $180^\circ$, with
\aif~$=\vert\phi_b^{Gd}\vert+\vert\phi_b^{Fe}\vert$, where
$\phi_b$ is the angle of the moment at the boundary layer.

The in-plane DW widths at the Gd/Fe interfaces have been
calculated as shown in Fig.~\ref{fig:DWs}. The orientation of the
moments as a function of the distance to the interface is fitted
to an exponential function of the form $\phi(t')=a \exp (b t')$,
where $t'$ is the distance to the interface, $a = \phi(t'=0)$ and
$b$ is a fit parameter. The DW width $\delta$ is the intersection
of the tangent of this function at $t'=0$ with the $\phi=0^\circ$
axis.

\begin{figure}[htb]
\begin{centering}
\includegraphics[bb=157 339 399 538,width=70mm]{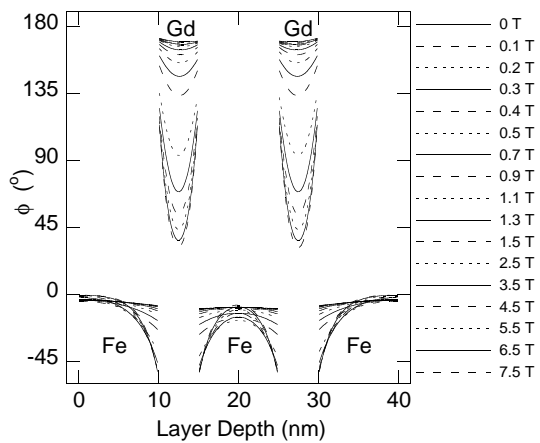}
\caption{{Angle $\phi$ against the layer depth $t$ for the
simulation with \tfe$~=10$~nm and \tgd$~=5$ nm. This is an example
of the Fe aligned state. For small H the magnetisation in the Fe
layer is fully aligned with the applied external field. At large
$H$ twisting occurs in the layers and \aif\ is reduced.}}
\label{fig:Gd5Fe10}
\end{centering}
\end{figure}

Three magnetic states have been identified in Gd/Fe
multilayers.\cite{Che91a,Duf93a,Hah95a,Mcg96a} Upon application of
a small external field $H$ either the Fe layer or the Gd layer
aligns with $H$. The ratio between the total magnetic moment per
element $M^{Gd}:M^{Fe}$ determines whether the Fe aligned (F) or
Gd aligned (G) states prevails. For instance, Hosoito \ea\ report
for a [Fe/Gd]$_{15}$ multilayer film that the G state is present
for low temperatures $T<120$ K where $M^{Gd}>M^{Fe}$, and the Fe
aligned state prevails for high temperatures $T>=140$ K, where
$M^{Gd}$ has become smaller than $M^{Fe}$ since \msgd\ is
significantly reduced.\cite{Hos02a} Both states are characterised
by \aif~$\sim 180^{\circ}$ in small fields. A typical example of a
micromagnetics simulation of the Fe aligned state is given in
Fig.~\ref{fig:Gd5Fe10}, showing a sample with \tfe$~=10$~nm and
\tgd$~=5$ nm. This simulation will be referred to as the \emph{F
state simulation} in the remainder of the article.

\begin{figure}[htb]
\begin{centering}
\includegraphics[bb=60 432 300 568,width=80mm]{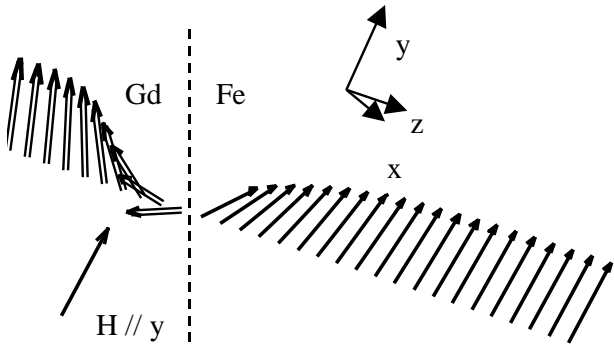}
\caption{Three dimensional view of the $\mu_0H=7.5$ T data of
Fig.~\protect{\ref{fig:Gd5Fe10}}. For clarity only the data are
plotted for $12.5<t<20$ nm, showing half of a Gd layer (thick
arrows) and half of an Fe layer (thin). The magnetic field is
parallel to the $y$-axis. At the interface, the Fe layer is
-45$^\circ$ off the $y$-axis and the Gd layer is +100$^\circ$ off
the $y$-axis.} \label{fig:spins}
\end{centering}
\end{figure}

Upon application of a large external field, most of each layer
will align parallel with the field. Nevertheless, $\phi$ will be
nonzero near the interfaces. This is called the twisted (T) state.
Fig.~\ref{fig:spins} sketches the alignment of the layers with the
applied field (parallel to the $y$-axis) and how the magnetisation
rotates away in the $xy$-plane near the interfaces to accommodate
the AFM coupling, thereby creating \ip\ DWs.

\begin{figure}[htb]
\begin{centering}
\includegraphics[bb=157 339 399 538,width=80mm]{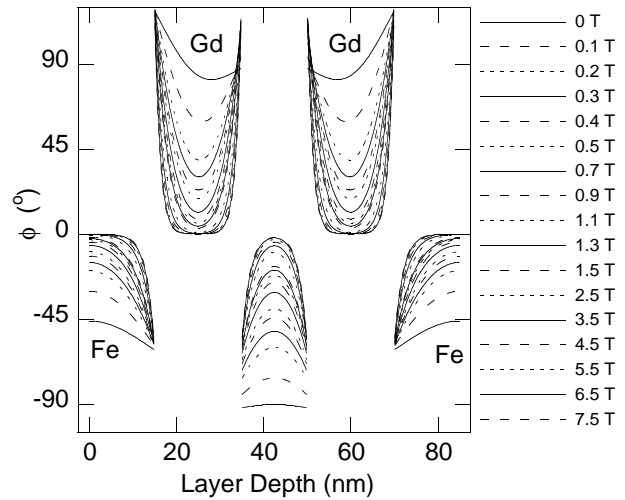}
\caption{{$\phi$ against $t$ for the simulation with \tfe$~=15$ nm
and \tgd$~=20$ nm. This simulation illustrates the twisted ground
state. Even at small $H$ \aif\ decreases rapidly with $H$ and DWs
are formed in both Fe and Gd layers. $\phi_c=0^\circ$ at
relatively small $H$ but $\phi_b\gg 0^\circ$ for all $H$. Note
that the twisting is more pronounced at low $H$ in the top and
bottom layer as recently reported by Haskel \emph{et al.}
\protect{\cite{Has03a}}}} \label{fig:Gd20Fe15}
\end{centering}
\end{figure}

The twisted state originates from the balance between the AFM
coupling at the interface (that tries to align one of the layers
antiparallel to $H$) and the Zeeman energy (which tries to align
all layers parallel to $H$). Fig.~\ref{fig:Gd20Fe15} shows the
\emph{twisted state simulation} in a sample with \tfe$~=15$ nm and
\tgd$~=20$ nm. The twisting of the FM layers is clearly present at
all $H>0$. Note that the layers at the edge of the samples rotate
faster, which is due to the free surface.\cite{Has03a} Also the
high field data in Fig.~\ref{fig:Gd5Fe10} shows the twisted phase.
The lower exchange coupling of Gd allows a larger angle between
neighbouring spins, \ie\ favouring narrower DWs and DWs
incorporating a larger twist, compared with the Fe layer.

\begin{figure}[htb]
\begin{centering}
\centering
\includegraphics[bb=184 311 414 559,width=70mm]{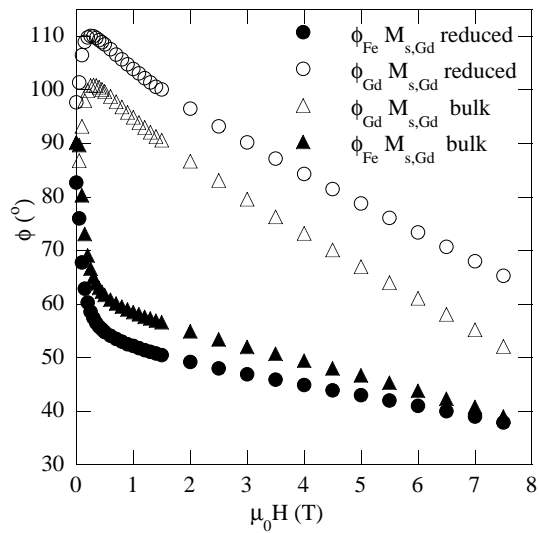}
\caption{{Angle at the boundary $\phi_b$ as a function of external
field. Open symbols correspond to $\phi_b^{Gd}$ and solid symbols
to $\phi_b^{Fe}$. Circles and triangles correspond to simulations
using \msgd = 1508 emu cm$^{-3}$ and \msgd = 2010 emu cm$^{-3}$,
respectively. }} \label{fig:all}
\end{centering}
\end{figure}

In experimental reports on Gd/Fe multilayers, the saturation
magnetisation of the thin Gd layers was found to be reduced by
20\% to 40\%.\cite{Ish99a,Hah95a,Hos02a} We have studied the
effect of reducing \msgd\ by comparing simulations of a \tfe\ =
\tgd\ = 30 nm multilayer. Two simulations have been performed, one
with \msgd\ = 2010 emu cm$^{-3}$  corresponding to the bulk value
and the other with \msgd\ = 1508 emu cm$^{-3}$ in the range of the
literature values found for thin Gd/Fe multilayers. The angles at
the interface $\phi_b$ for both Gd and Fe layers, are extracted
from the simulation and plotted in Fig.~\ref{fig:all}.

Changing \msgd\ to 75\% of the bulk value has only minor effects
on the simulation. Experimental results indicate that the
saturation magnetisation in thin Gd films might be reduced by 20\%
to 40\%. Therefore \msgd\ was reduced to 75\% of the bulk value in
all other simulations.

We have also simulated a multilayer with a reduced interface
coupling $A_{int}=-A_{Gd}=-0.75~\mu$erg cm$^{-1}$. The main
features, relative to the "standard" simulation with
$A_{int}=-1.0~\mu$erg cm$^{-1}$, were strongly reduced \aif\ and
boundary angles but slightly higher DW widths for all fields.
However the DW width actually increased slightly with increasing
field at high fields. This is related to the drastic decrease of
\aif\ and the very small twists in the DWs at those fields.

Generally speaking the stronger the AFM coupling, the larger the
interface angle and boundary angles. Therefore, the DWs will
become narrower and have a larger twist. The narrower DWs will
have less overlap and the twisted phase will still be present at
smaller layer thicknesses.

\begin{figure}[htb]
\begin{centering}
\includegraphics[bb=19 567 262 820,width=80mm]{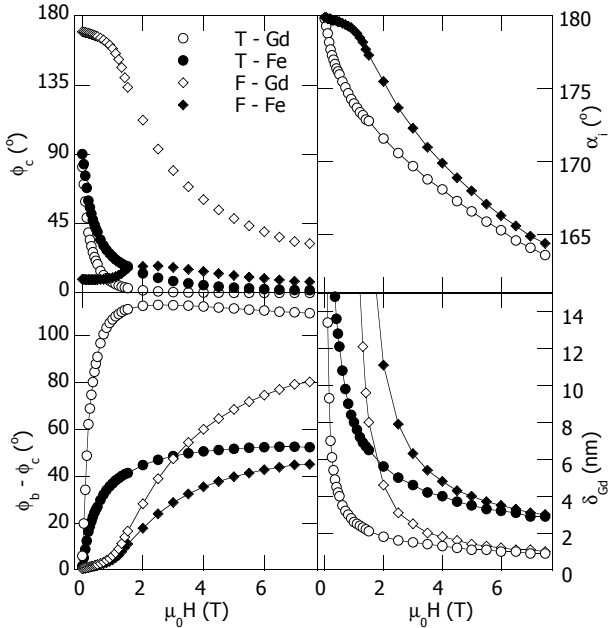}
\caption{{Data extracted of the twisted state simulation with
\tfe\ = 15 nm and \tgd\ = 20 nm and the Fe aligned state
simulation with \tfe\ = 10 nm and \tgd\ = 5 nm. The angle of the
magnetic moment at the centre $\phi_c$, the interface angle \aif,
the twist in the DW $\phi_b-\phi_c$ and the width of the DW
$\delta$ are plotted vs the applied field $H$. The figure
highlights the different behaviour of the Fe and the Gd layer.
Open circles (T - Gd) correspond to the Gd layer of the twisted
state simulation; Closed circles (T - Fe) correspond to the Fe
layer of the twisted state simulation. Open diamonds (F - Gd)
represent the Gd layer of the Fe aligned state simulation and
closed diamonds (F - Fe) represent the Fe layer of the Fe aligned
state simulation.}} \label{fig:FeGd5_30}
\end{centering}
\end{figure}

We now look at the effect of sweeping the magnetic field on the
magnetic state of the Gd/Fe multilayers. We have plotted the
angles $\phi_c$, \aif, $\phi_b-\phi_c$ and DW width $\delta$ as a
function of $H$ in Fig.~\ref{fig:FeGd5_30}, data is taken from the
Fe aligned state simulation (Fig.~\ref{fig:Gd5Fe10}) and twisted
state simulation (Fig.~\ref{fig:Gd20Fe15}). The main difference
between the aligned states and the twisted state can be seen in
the top-right panel of Fig.~\ref{fig:FeGd5_30}. In the twisted
state simulation \aif\ decreases rapidly with $H$, whereas in the
Fe aligned state simulation \aif~$\approx 180^{\circ}$ for small
fields. The reduction of \aif\ with increasing field is in
agreement with the experimental results reported on anti phase
boundaries in Fe$_3$O$_4$ \cite{Eer02b} and AFM coupled interfaces
in Gd/Fe multilayers .\cite{Pri04a}

The decrease of \aif\ in the twisted state simulation is
accompanied by the formation of DWs at the interfaces of the thick
layers. In the Fe aligned state simulation  \aif, $\phi_b$ and
$\phi_c$ are only weakly dependent on the applied magnetic field.
Each Gd or Fe layer remains almost perfectly aligned, \ie\ no
twisting in the layer. At about $\mu_0H=0.7$ T, there is a change
in the gradient of \aif, $\phi_b$ and $\phi_c$ with $H$. The
layers also start to twist and form wide DWs.

In both simulations we find that with increasing field the DWs
become narrower and \aif\ decreases. The curves for \dgd\ converge
at $\mu_oH\sim5$ T. Apparently, the Gd DWs in both simulations
become independent of \tgd\ at these high fields. The data for
\dfe\ suggest that the same will happen in the Fe layers at even
higher fields ($\mu_0H>8$ T).

\begin{figure}[htb]
\begin{centering}
\includegraphics[bb=20 568 265 820,width=80mm]{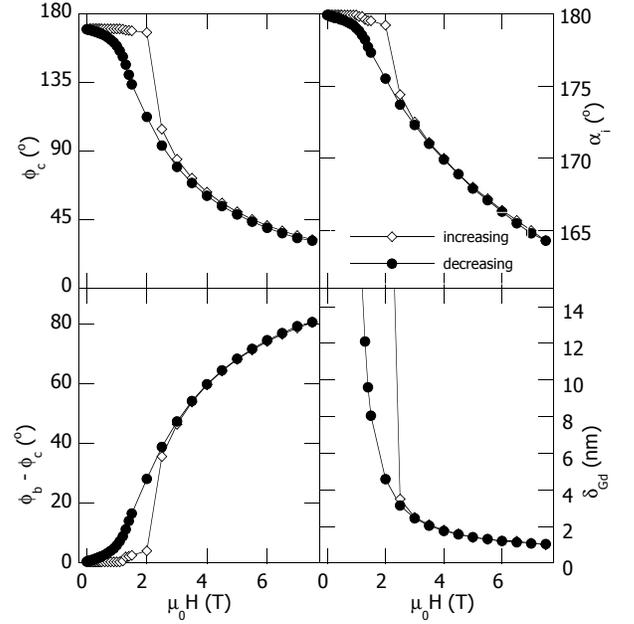}
\caption{{Data extracted of the F aligned simulation with \tfe\ =
10 nm and \tgd\ = 5 nm. Data has been taken in decreasing field,
followed by increasing field. Note the hysteretic behaviour at
small fields. Y-axis symbols are the same as in
Fig.~\ref{fig:FeGd5_30}}} \label{fig:F}
\end{centering}
\end{figure}

The samples in Fig.~\ref{fig:FeGd5_30} have been simulated whilst
sweeping the field from $\mu_0 H=7.5$~T to $\mu_0 H=0$ T. In
Fig.~\ref{fig:F}, also the increasing field sweep is shown for the
Fe aligned state simulation with \tfe\ = 10 nm and \tgd\ = 5 nm,
taking the magnetic configuration at $H=0$ as starting
configuration.

Looking at the behaviour at small fields, during the decreasing
field sweep, the Fe (Gd) layers rotate slowly but steadily to be
parallel (antiparallel) with the field direction at $\mu_0H=0$ T.
In the increasing field sweep the layers initially stay parallel
just as they are at $\mu_0 H=0$ T. Between $\mu_0 H=2$ T and
$\mu_0 H=2.5$ T, the magnetic configuration abruptly changed to
coincide with the decreasing field curve. Clearly the increasing
field simulation got trapped in a metastable state, whereas the
decreasing field sweep changed smoothly with $H$.

The simulation for the twisted state sample (with \tfe$~=15$ nm
and \tgd$~=20$ nm) shows no significant difference between the
decreasing and increasing field sweeps. In a second twisted state
simulation, after the $H\parallel y$ decreasing field sweep, the
increasing field is parallel to $x$. $H$ is then parallel with
either the Fe or the Gd magnetisation. During the increasing field
sweep a similar hysteresis like transition is observed as in the
Fe aligned state simulation. This demonstrates the importance of
the initial magnetic state and the direction of the field sweep.

\begin{figure}[htb]
\begin{centering}
\includegraphics[bb=20 568 265 820,width=80mm]{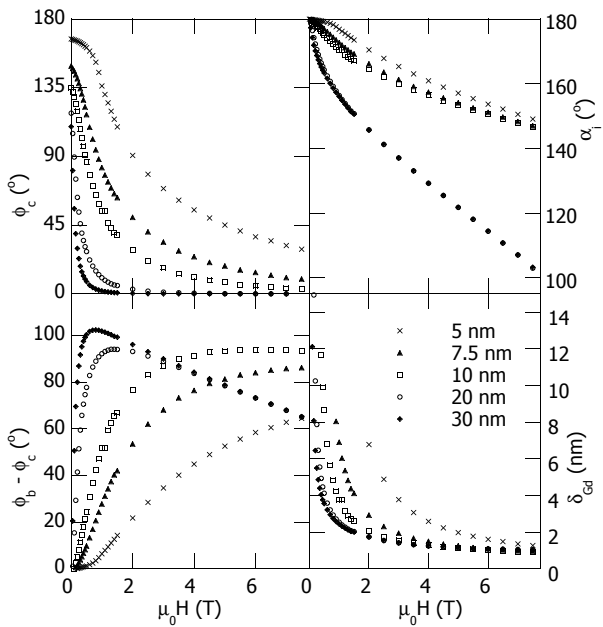}
\caption{{The simulation of Gd/Fe multilayers for various \tfe\ =
\tgd. The simulations for \tfe\ = \tgd\ = 20 nm and \tfe\ = \tgd\
= 30 nm are identical above $\mu_0H=2$ T. For clarity only the the
data for the Gd layers are shown. Y-axis symbols are the same as
in Fig.~\ref{fig:FeGd5_30}}} \label{fig:constant}
\end{centering}
\end{figure}

We have shown the typical simulations for the twisted state (\tfe\
= 15 nm and \tgd\ = 20 nm) and the Fe aligned state (\tfe\ = 10 nm
and \tgd\ = 5 nm). We now investigate the effect of changing the
layer thicknesses whilst keeping the ratio $M^{Fe}:M^{Gd}$
constant. If the magnetic configuration is only controlled by this
ratio, then these simulations should exhibit qualitatively similar
behaviour. In Fig.~\ref{fig:constant} we plotted as a function of
$\mu_0H$ the data extracted for different values of \tfe = \tgd.

At low fields and \tfe\ = \tgd\ $<10$ nm the F phase prevails, as
expected since $M^{Fe}>M^{Gd}$ in these \tfe\ = \tgd\ multilayers.
The prevalence of the F phase can be deduced from the small value
of $\phi_b-\phi_c$, the slow decrease of \aif, and the large value
of \fcgd$\gg90^\circ$. Only for higher fields a transition to the
twisted phase takes place and \aif\ starts to decrease fast. These
simulations with \tfe\ = \tgd\ $<10$ nm are similar to the Fe
aligned state simulation shown in Fig.~\ref{fig:Gd5Fe10}. However,
for larger \tfe\ = \tgd\ the multilayer adopts the twisted phase
at all fields even though $M^{Fe}>M^{Gd}$. As in the twisted state
simulation (\tfe$~=15$ nm and \tgd$~=20$ nm) \aif\ is much smaller
than $180^\circ$ at any $\mu_0H>0$ T and $\phi_b\gg\phi_c$ for
small fields.

\begin{figure}[htb]
\begin{centering}
\includegraphics[bb=95 405 330 613,width=75mm]{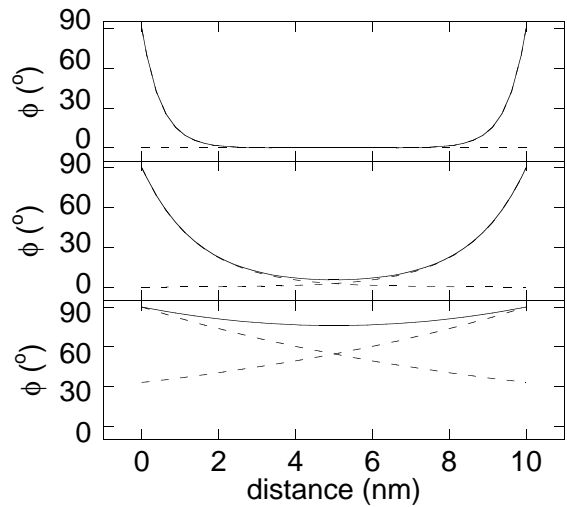}
\caption{{Sketch of the DW overlap. The DW width is reduced from
the top to the bottom panel. The surface tension of the
overlapping DWs enhances the reduction of the twist.}}
\label{fig:DWoverlap}
\end{centering}
\end{figure}

The different behaviour for small and large \tfe\ = \tgd\ can be
explained by the overlap of DWs in the centre of each layer. For
high fields 2\dgd~$<<$~\tgd\ and the two DWs do not meet at the
centre of the layer; they act independently. However at smaller
fields (larger \dgd) and smaller \tgd\ the DWs do overlap. For
instance, Fig.~\ref{fig:constant} shows that at $\mu_0H=2.5$ T the
Gd DW is 4.8 nm wide for the multilayer with \tfe\ = \tgd\ = 5 nm.
The overlapping DWs reduce the twist by means of the surface
tension\cite{Chikazumi} in the DW and the thin layers each become
FM aligned. In Fig.~\ref{fig:DWoverlap} this scenario has been
sketched. Since $M^{Fe}>M^{Gd}$ the Fe layer will align parallel
to the applied field, and \fcgd\ will approach $180^\circ$ at
$H=0$, resulting in the F phase. For thicker layers, the twist is
removed, \ie\ the layers become totally FM aligned, only when the
field is very small during a decreasing magnetic field sweep.
These small fields are insufficient to rotate the Fe layers
parallel with the applied field. Consequently, both layers are
perpendicular to the applied field direction, $\phi_c\sim \pm
90^\circ$ at $H=0$ and the twisted phase is the ground state for
thicker layers.

The DW thickness depends on the layer thickness as can be observed
in the bottom right corner of Fig.~\ref{fig:constant}. \dgd\
increases with decreasing layer thickness at constant field. We
also note that with decreasing layer thicknesses, the $H-$\dgd\
curve converges at larger fields. This confirms the independent
behaviour of narrow DWs in thicker layers as discussed above.

\begin{figure}[htb]
\begin{centering}
\includegraphics[bb=37 565 288 765,width=75mm]{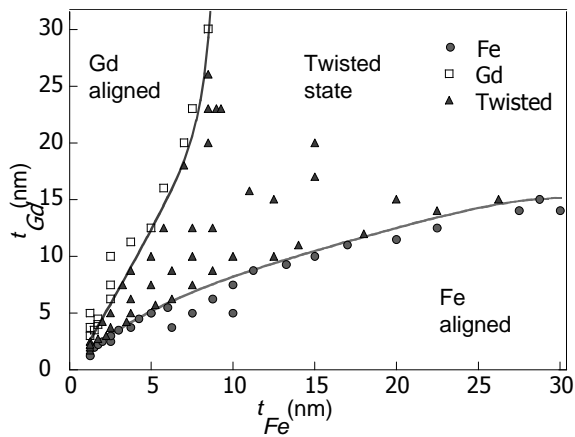}
\caption{{Phase diagram as a function of the total Fe
magnetisation (bottom axis) and the total Gd magnetisation (left
axis). The drawn lines indicate the boundaries between the Gd
aligned (open squares), the twisted (grey triangles) and the Fe
aligned (grey circles) states. The corresponding layer thicknesses
are given on the top (\tfe) and right (\tgd) axes.}}
\label{fig:phasediagram}
\end{centering}
\end{figure}

We have shown that there is a transition from the F phase at \tfe\
= \tgd\ = 5 nm to the T phase at \tfe\ = \tgd\ = 30 nm, even
though the ratio $M^{Fe}:M^{Gd}$ remained constant. Therefore, the
magnetic ground state does not only depend on the ratio between
the total magnetisation but also on the individual layer
thicknesses. We have investigated this dependency for a range of
\tfe's and \tgd's. We have chosen to use a single variable as
indicator for the magnetic ground state: \fcgd\ at $H=0$. If
\fcgd\ is between $45^\circ$ and $135^\circ$ the ground state of
the simulation is designated as the T phase. Outside these
boundaries either the F phase (\fcgd~$>135^\circ$) or the G phase
(\fcgd~$<45^\circ$) is assigned. The phase diagram of the magnetic
ground state as a function of \tfe\ and \tgd\ is plotted in
Fig.~\ref{fig:phasediagram}.

The phase diagram confirms that the boundaries between the three
magnetic ground states are not at fixed ratios of $M^{Fe}:M^{Gd}$.
The data strongly suggests that the G phase is absent if \tfe\
$>9$ nm. One expects the G phase for multilayers with $M^{Gd}\gg
M^{Fe}$ but the phase diagram indicates that this is only the case
if \tfe~$<9$ nm. Above this value, no matter how thick the Gd
layer is, the simulation will always yield the T phase. Similarly,
the F phase appears to be absent if \tgd\ exceeds $\approx 16$ nm.
We can conclude from these observations that when both layer
thicknesses exceed these critical values only the T phase is
present, independent on the ratio $M^{fe}:M^{Gd}$.

Below $t_{tot}=20$ nm, the fraction in the T phase drops rapidly
with decreasing $t_{tot}$. Already at high fields, the DWs overlap
and reduce their twist. These thin layers do not allow the
existence of DWs at small fields. This suppresses the T phase for
ultra thin multilayers.

In conclusion, using micromagnetics Gd/Fe multilayers consisting
of three Fe layers and two Gd layers have been modelled to study
the effect of \tfe\ and \tgd\ on the magnetic ground state. The
twisted phase can be distinguished from the Fe and Gd aligned
phases by the rapid decrease of \aif\ at small fields. At the same
time wide DWs are formed at the edges of the thick layers. In both
aligned states \aif\ remains almost 180$^\circ$ in small fields.
Each Gd or Fe layer remains almost perfectly aligned, \ie\ no
twisting in the layer. Above a threshold field, the aligned phases
are transformed into the twisted phase. In all cases, at large
fields the exchange constants keep \dfe~$>$~\dgd, with the twist
in the Gd DW larger than in the Fe DW.

When the magnetic field is applied parallel with the magnetisation
of one of the layers in an increasing field sweep hysteretic
behaviour can be observed. The sample will show a near perfect
alignment of that layer with the field and a sharp transition to
the twisted state is seen.

We have found that the magnetic ground state depends both on the
ratio between the total magnetisation for Fe and Gd and on the
individual layer thicknesses. The different behaviour for small
and large layer thicknesses can be explained by the amount of
overlap of the DWs.

The phase diagram shows that independent on the ratio
$M^{Fe}:M^{Gd}$ only the twisted phase is allowed when both layer
thicknesses exceed their respective critical values. For thin
layers all the magnetic ground states are allowed and the ground
state found depends on the ratio $M^{Fe}:M^{Gd}$. Ultra thin
layers do not allow the existences of DWs at small fields. This
suppresses the twisted phase.

This work was supported by an E.U. Marie Curie Fellowship, the UK
EPSRC and the Royal Society.

\bibliographystyle{e:/texmf/data/apsrevbas4subm}
\bibliography{e:/texmf/data/Bas}

\end{document}